\documentclass[%
 reprint,
 amsmath,amssymb,
 aps,
]{revtex4-1}

\usepackage{graphicx}
\usepackage{dcolumn}
\usepackage{bbm}
\usepackage{bm}
\usepackage{geometry}
\usepackage{braket}
\usepackage{lipsum}
\usepackage{mathtools}
\usepackage{lipsum}
\usepackage{xcolor}
\usepackage[colorlinks=true,allcolors=blue]{hyperref}
\newcommand*{\SavedEqref}{}
\let\SavedEqref\eqref
\renewcommand*{\eqref}[1]{%
	\begingroup
	\hypersetup{
		linkcolor=linkequation,
		linkbordercolor=linkequation,
	}%
	\SavedEqref{#1}%
	\endgroup
}

\begin{document}


\title{Generalized su(1,1) algebra and the construction of nonlinear coherent states for  P\"{o}schl-Teller potential}

\author{Abdessamad Belfakir}
\email{abdobelfakir01@gmail.com}
\affiliation{Equipe des Sciences de la Mati\`ere et du Rayonnement(ESMAR),\\
	Faculty of Sciences, Mohammed V University,
	Av. Ibn Battouta, B.P. 1014, Agdal,
	Rabat, Morocco}
		\author{Yassine Hassouni}
		\email{yassine.hassouni@gmail.com}
\affiliation{Equipe des Sciences de la Mati\`ere et du Rayonnement(ESMAR),\\
		 Faculty of Sciences, Mohammed V University,
		Av. Ibn Battouta, B.P. 1014, Agdal,
		Rabat, Morocco}



%





\begin{abstract}
	\section*{abstract}
 We introduce a generalization structure of the su(1,1) algebra  which depends on a function of one generator of the algebra, $f(H)$. Following the same ideas developed to the generalized  Heisenberg algebra (GHA) and to the generalized su(2), we show that a symmetry is present in the sequence of eigenvalues of one generator of the algebra. Then, we construct the Barut-Girardello coherent states associated with the generalized su(1,1) algebra for a particle in a P\"oschl-Teller potential. Furthermore, we compare the time evolution of the uncertainty relation of the constructed coherent states with that of GHA  coherent states. The generalized su(1,1)  coherent states are very localized.
\end{abstract}

\pacs{Valid PACS appear here}
\maketitle

  {\quotation\noindent{\bf Keywords:}  Generalized Heisenberg algebra (GHA), su(1,1) algebra, Coherent state, P\"oschl-Teller potential.
  	
  	\endquotation} 
\section{Introduction}
Coherent states of the harmonic oscillator introduced first by Schr\"{o}dinger in 1926 \cite{shrodinger} are the only quantum states that minimize the Heisenberg uncertainty relation for the position and momentum operators \cite{shrodinger,Glauber}. The dynamic of expectation values of  position and momentum operators on these states has the same form as that of their classical counterpart, for this reason, they are called quasi-classical states\cite{PhysRevLett.10.277}. More recently, in 1960s Glauber \cite{Glauber},  Sudarshan \cite{PhysRevLett.10.277} and Klauder \cite{Klauder1} studied  these states widely in quantum optics showing their important physical applicability. Then, coherent states have been constructed for physical systems other than the harmonic oscillator such as a free particle in a square well potential \cite{Hassouni2}, hydrogen atom \cite{hydrogenatom,HydroatomCS} and  P\"{o}schl-Teller potential \cite{Curado2}. These states were called  nonlinear coherent states. Additionally, several generalizations of coherent states have been introduced \cite{Hassouni2,Klauder3,Roknizadeh,PhysRevA.54.4560,Gazeau} and different approaches for constructing them  have been investigated such as Perelomov and Klauder-Gilmore approaches \cite{Perelomov,Klau}. Subsequently, coherent states and squeezed states can be algebraically constructed by using a unified approach proposed in \cite{PhysRevLett.72.1447}.\\In  recent  years, the generalized Heisenberg algebra (GHA) \cite{Monteiro1,Monteiro2,Hassouni} has been constructed and  more recently it was considered as an alternative method to construct new solvable models from old ones\cite{Bagarello1}. By knowing the spectrum of the physical system, the characteristic function of the GHA connecting between two successive dimensionless energy eigenvalues can be derived and the associated algebra can be constructed \cite{Monteiro1,Monteiro2,Hassouni}. The GHA has been applied to several physical systems  such as  P\"{o}schl-Teller potential \cite{Curado2,Bagarello1} and Morse potential \cite{Angelova1}. Further, the nonlinear coherent states associated with GHA, called the GHA coherent states, have been constructed for these physical systems \cite{Hassouni2,Curado2,MorseCS}.\\
In this paper, we will introduce  a generalization structure of the su(1,1) algebra and show that a hidden symmetry is present in the sequence of eigenvalues of one algebra generator similarly to the  GHA and to the generalized su(2) algebra constructed in \cite{Curado00,Curado01}. We apply the introduced generalized su(1,1) algebra to the P\"{o}schl-Teller potential 
and show that it can be applied to practically   other physical systems  exactly as the GHA. For a particle in a P\"{o}schl-Teller potential, we construct the coherent states associated with the generalized su(1,1) algebra and compare their behavior with that of GHA coherent states \cite{Curado2}. The generalized su(1,1) coherent states constructed here maintain an important localizability in the time compared with GHA coherent states.\\
This paper is structured as follows: in section (\ref{section2}) we recall the GHA and introduce the generalized su(1,1) algebra. Then, we construct the associated generalized harmonic oscillator ladder
operators.  In section (\ref{section3}) we apply the generalized su(1,1) algebra to the P\"{o}schl-Teller potential and find the physical realizations of the algebra generators in terms of position and differential operators. Moreover, In section (\ref{section4}) we shall construct the  generalized su(1,1) Barut-Girardello  nonlinear coherent states for the P\"{o}schl-Teller potential and compare the behavior of  the time evolution of the uncertainty relation of canonically conjugate operators for both nonlinear coherent states. Finally, our conclusions are given in (\ref{section5}) .
 		\section{  Generalized su(1,1) algebra }\label{section2}
 			\subsection{A review on GHA}
The  GHA  has been constructed and applied to several physical systems\cite{Monteiro1,Monteiro2,Hassouni}.  This algebra is described by three operators $H$, $A$ and $A^\dagger$ satisfying the following relations 
 	\begin{equation}\label{GHA1}
 HA^\dagger=A^\dagger f(H),
 	\end{equation}
 	\begin{equation}\label{GHA2}
 AH=f(H)A ,
 \end{equation}
 	and 
 	\begin{equation}\label{GHA3}
[A,A^\dagger]=AA^\dagger-A^\dagger A= f(H)-H,
 \end{equation}
 	where $A=(A^{\dagger})^\dagger$, $H$ is the dimensionless Hamiltonian of the physical system under consideration and $f(H)$ is an analytical function of $H$, called the characteristic function of the algebra. A large class of Heisenberg algebras can be obtained by particular choices of the analytical function $f$. Particularly, if $f(H)=H+1$ and $H$ is the dimensionless Hamiltonian of the harmonic oscillator, the GHA defined in (\ref{GHA1})-(\ref{GHA3}) recover the ordinary Heisenberg algebra spanned by $H$ and the creation and  annihilation operators of the harmonic oscillator \cite{Hassouni}.\\
 	 The Casimir operator of the GHA is given by
 	\begin{equation}\label{Casimir}
 \Gamma=A^\dagger A-H=AA^\dagger -f(H).
 	\end{equation}
 The irreducible representation of the GHA is given through the eigenvectors $\ket{n}$ of the Hamiltonian $H$, $H\ket{n}=\varepsilon_n\ket{n}$ such that 
 \begin{equation}
\varepsilon_{n+1}=f(\varepsilon_{n}),\quad n=0,1,2,\dots,
 \end{equation}
 where $\varepsilon_{n+1}$, $\varepsilon_{n}$ are two successive energy eigenvalues. Then, the Fock space representation of the GHA is given through the eigenvalue $\varepsilon_{0}$ corresponding to the ground state $\ket{0}$ and an eigenvalue $\varepsilon_{n}$ is the $n$-iterate of $\varepsilon_{0}$ under $f$, i.e, 
 \begin{equation}
 \varepsilon_{n}=f^n(\varepsilon_{0}).
 \end{equation}
 Assuming that $A\ket{0}=0$ and by using $(\ref{GHA1})-(\ref{Casimir})$, we can show that 
 \begin{equation}
A^\dagger\ket{n}=N_n\ket{n+1},
 \end{equation}
  \begin{equation}
 A\ket{n}=N_{n-1}\ket{n-1},
 \end{equation}
 where 
 \begin{equation}
N_n^2=\varepsilon_{n+1}-\varepsilon_{0}.
 \end{equation}
 The operators $A^\dagger$, $A$ are then the  creation and the annihilation operators of GHA, respectively.
 \subsection{Generalized su(1,1) algebra}
 Let us now introduce an algebraic structure that generalizes the su(1,1) algebra and give the associated Fock space representation. Similarly to the GHA, we will show that the generalized su(1,1) algebra can be constructed once the spectrum of the physical system under consideration is known, i.e., once the characteristic function of the algebra is determined. Let $H$, $J_+$ and $J_-$ be three operators obeying the following relations 
 \begin{equation}\label{j+}
 H J_{+}= J_{+}f(H),
 \end{equation}
 \begin{equation}\label{j-}
 J_{-}H=f(H)J_{-},
 \end{equation}
 and 
 \begin{equation}\label{H}
 [J_{+},J_{-}]=(H-f(H))(H+f(H)-1),
 \end{equation}
 where $J_{-}=J_{+}^\dagger$, $H$ is the dimensionless Hamiltonian of the system under consideration and $f(H)$ is a given function of $H$. In fact, $H$ can be any hermitian operator.\\
 For the particular case, $f(H)=H+1$, the relations (\ref{j+})-(\ref{H}) become
 \begin{equation}
 [H,J_{+}]=J_{+},
 \end{equation}
 \begin{equation}
  [H,J_{-}]=-J_{-},
 \end{equation}
 and 
 \begin{equation}
 [J_{+},J_{-}]=-2H.
 \end{equation}
These  relations are the well-known su(1,1) algebra. Thus, the relations (\ref{j+})-(\ref{H}) contains the su(1,1) algebra by choosing the specific function $f(H)=H+1$, as a particular case. For this reason we call it the generalized su(1,1) algebra. Let us consider the operator $C$ given by
\begin{equation}\label{cas}
C=J_+J_{-}-H(H-1)=J_{-}J_+-f(H)(f(H)-1).
\end{equation}
 By using  (\ref{j+})-(\ref{H}), we can show that $C$ satisfies the following commutation relations
\begin{equation}
[C,H]=[C,J_{\pm}]=0,
\end{equation}
showing that $C$ is the Casimir operator of the algebra (\ref{j+})-(\ref{H}). Now, let us provide the Fock space representation of the generalized su(1,1) algebra introduced in (\ref{j+})-(\ref{H}). Let $\ket{n}$ be an eigenvector of $H$, $H\ket{n}=\varepsilon_{n}\ket{n} $ and let us assume that we have an infinite irreducible representation ($n=0,1,\dots.$). Applying (\ref{j+}) on $\left\vert n \right\rangle$, we find that
	\begin{equation}
	H(J_{+}\left\vert n\right\rangle)=J_{+}f(H)\left\vert n\right\rangle=f(\varepsilon_{n})(J_{+}\left\vert n\right\rangle).
	\end{equation}
	Then, $J_{+}\left\vert n\right\rangle$ is an  eigenvector of $H$ with eigenvalue $f(\varepsilon_{n})$. Let $\varepsilon_{n+1}=f(\varepsilon_{n})$, Thus $J_{+}\left\vert n\right\rangle$ is proportional to $\left\vert n+1\right\rangle$ showing that  $J_{+}$ is a raising operator.
	\\Following the same procedure for $J_{-}$, from (\ref{j-}) we have  
	\begin{equation}
	J_{-}H\left\vert n\right\rangle=f(H)J_{-}\left\vert n\right\rangle=\varepsilon_{n}(J_{-}\left\vert n\right\rangle),
	\end{equation}
	which implies that $J_{-}\left\vert n\right\rangle$ is an eigenvector of $H$ with eigenvalue $\varepsilon_{n-1}$, implying that $J_{-}\left\vert n\right\rangle$  is proportional to $\left\vert n-1\right\rangle $. Thus, $J_{-}$ is an annihilation operator. By applying the Casimir (\ref{cas}) on $\ket{0}$ and assuming that $J_-\ket{0}=0$, we can show that 
\begin{equation}\label{j++}
J_+\ket{n}=\sqrt{(\varepsilon_{n+1}-\varepsilon_{0})(\varepsilon_{n+1}+\varepsilon_{0}-1)}\ket{n+1},
\end{equation}
\begin{equation}\label{j--}
J_-\ket{n}=\sqrt{(\varepsilon_{n}-\varepsilon_{0})(\varepsilon_{n}+\varepsilon_{0}-1)}\ket{n-1},
\end{equation}
for $n=0,1,\dots.$ The operators $J_+$, $J_-$ are then the ladder operators of the generalized su(1,1) algebra. We note that $J_-\ket{0}=0$, which means that the vacuum state condition is verified. The first two  relations defining the generalized su(1,1) algebra (\ref{j+})-(\ref{j-}) have the same form as those of the GHA (\ref{GHA1})-(\ref{GHA2}). Furthermore, the GHA and the algebra given in (\ref{j+})-(\ref{H}) have the same characteristic function relying two successive dimensionless energy eigenvalues $\varepsilon_{n+1}$ and $\varepsilon_{n}$ of the physical system under consideration. Moreover, since any quantum system having $\varepsilon_{n+1}=f(\varepsilon_{n})$ can be perfectly described by GHA, this system can also be described by the generalized su(1,1) algebra.\\
A deformation of the su(1,1) algebra by two analytical functions of an hermitian operator has been constructed in \cite{Delbecq_1993}. This  deformed algebra contains the su(1,1) algebra as a  particular case and the approach can be applied to particular problems depending on the functions of deformations which can be linear or nonlinear. Our approach differs from the one given in \cite{Delbecq_1993} since it can be applied to  physical systems whose spectrum is known as the GHA approach. Another particularity of the introduced generalized su(1,1) algebra (\ref{j+})-(\ref{H}) compared with those given in \cite{Delbecq_1993} is that here, the corresponding function of deformation is the function relying between two eigenvalues of one generator of the algebra. Another nonlinear deformed su(1,1) algebras were applied to the P\"{o}schl-Teller potential \cite{Quesne_1999,Chen_1998}. However, these approaches are complicated and can not be extended to other physical systems. Here, we have shown that physical systems can be described algebraically in an easier way once the characteristic function is determined. In section (\ref{section2}), we will apply the generalized su(1,1) algebra introduced in this work to the P\"{o}schl-Teller system and show how this system can be simply described by this algebra. Another motivation of introducing this algebra is the construction of another type of nonlinear  coherent states for physical systems. Let us note that the difference between our generalized su(1,1) algebra and the deformed su(1,1) algebras already existed in the literature \cite{Quesne_1999,Chen_1998} is exactly the difference between the GHA and the Generalized deformed oscillator algebras introduced in \cite{Daskaloyannis_1991}. In the first, the function of deformation appears in the spectrum such that $f(\varepsilon_n)=\varepsilon_{n+1}$. However, both nonlinear algebras contains the harmonic oscillator algebra as a particular case. In \cite{Berrada}, Berrada  \textit{et al.} introduced a deformed su(1,1) algebras and constructed the corresponding coherent states. However, the algebras introduced in \cite{Berrada} can not be applied to  physical system and the constructed coherent states are abstract and not related to physical systems. Here, the function of deformation is related to  systems and can be simply derived once the spectrum is known. Hence, the generalized su(1,1) constructed here can be applied to  physical systems having the property $\varepsilon_{n+1}=f(\varepsilon_n)$ and this function is the characteristic function of the algebra.
\subsection{Generalized su(1,1) algebra and generalized harmonic oscillator ladder operators}
We consider a general system whose generalized su(1,1) algebra is constructed and $H$ is its Hamiltonian and let $N$ be the corresponding number particle operator $N$, i.e., $N\ket{n}=n\ket{n}$ for $n=0,1,2,\dots$.  As already provided for GHA in \cite{Curado2}, we  define now an operator $B$ associated with the generalized su(1,1) algebra, which with its conjugate and $N$ satisfy the algebraic relations satisfied by the creation and annihilation operators $a^\dagger$, $a$ and $N$ of the harmonic oscillator. The operator $B$ is defined by 
\begin{equation}\label{B}
B=\sqrt{\dfrac{N+1}{(f(H)-\varepsilon_{0})(f(H)+\varepsilon_{0}-1)}}J_-,
\end{equation}
and the operator $B^\dagger$ is the adjoin of $B$. The operators inside the square root in (\ref{B}) are  hermitian. Then, their  square root  is well known. The action  of $B$ on a vector $\ket{n}$ is 
\begin{equation}
B\ket{n}=\sqrt{n}\ket{n-1},\quad n=0,1,2,\dots.
\end{equation}
It is then simple to prove that 
\begin{equation}
	[B,B^+]=\mathbbm{1},
\end{equation}
\begin{equation}
[N,B]=-B,
\end{equation}
\begin{equation}
[N,B^\dagger]=B^\dagger,
\end{equation}
where $\mathbbm{1}$ is the identity operator. Now, let us  introduce the canonically conjugate position-like and momentum-like operators $(X,P)$ associated with the generalized su(1,1) algebra. They are given by 
\begin{equation}\label{X}
X=\dfrac{L}{\sqrt{2}}(B+B^\dagger),
\end{equation}
\begin{equation}\label{PX}
P=\dfrac{i\hbar}{\sqrt{2}L}(B^\dagger-B).
\end{equation}
where $L$ is a constant having the dimension of length. From (\ref{X})-(\ref{PX}), we can show that 
\begin{equation}
[X,P]=i\hbar\mathbbm{1}.
\end{equation}
	\section{  Generalized su(1,1) algebra  and  P\"{o}schl-Teller potential}\label{section3}
	The one-dimensional P\"{o}schl-Teller (PT) is a Model having important applications in  physics \cite{doi:10.1080/00268979000101331,PhysRevB.72.115340}. The associated Schr\"{o}dinger equation is
	\begin{equation}\label{PT}
	\mathcal{H}\psi_n(x)=\left(-\dfrac{\hbar^2}{2m}\dfrac{d^2}{dx^2}+V_{PT}(x)\right)\psi_n(x)=E_n\psi_n(x),
	\end{equation}
	where
	\begin{equation}
	V_{PT}(x)=\dfrac{\hbar^2\pi^2}{2mL^2}\dfrac{\nu(\nu+1)}{\sin^2(\pi x/L)},
	\end{equation} 
such that $\nu\geq 0$, $0<x<L$ and $m$ is the mass of the particle in the  potential $V_{PT}(x)$. For $x\leq 0$ and $x\geq L$, the potential is infinite.\\
The discrete spectrum of the system is 
\begin{equation}\label{EPT}
E_n=\dfrac{\hbar^2\pi^2}{2mL^2}(n+\nu+1)^2,\quad n=0,1,2,\dots .
\end{equation}
The associated energy eigenfunctions  are given by 
\begin{equation}\label{psipt}
\psi_{n}(x)=c_{n}(\nu)\sin^{\nu+1}(\pi x/L)C_{n}^{\nu+1}(\cos(\pi x/L)),
\end{equation}
where 
\begin{equation}
c_{n}(\nu)=\Gamma(\nu+1)\dfrac{2^{\nu+1/2}}{\sqrt{L}}\sqrt{\dfrac{n!(n+\nu+1)}{\Gamma(n+2\nu+2)}},
\end{equation}
is the normalization constant, $\Gamma$ is the gamma function and $C_{n}^{\nu+1}$ are the Gegenbauer polynomials \cite{Wilhelm}.
From (\ref{EPT}), the spectrum of the dimensionless Hamiltonian $H=\mathcal{H}\dfrac{2mL^2}{\hbar^2\pi^2}$ is 
\begin{equation}\label{varepsi}
\varepsilon_n=(n+\nu+1)^2,\quad n=0,1,2\dots.
\end{equation}
Thus, the characteristic function of the generalized  su(1,1) algebra can be easily obtained. We have 
\begin{equation}
\varepsilon_{n+1}=(n+\nu+2)^2=(\sqrt{\varepsilon_{n}}+1)^2.
\end{equation} 
It follows that 
\begin{equation}
f(x)=(\sqrt{x}+1)^2.
\end{equation}
Consequently, The generalized su(1,1) algebra associated with the physical system is 
\begin{equation}\label{PT1}
[H,J_+]=2J_+\sqrt{H}+J_+,
\end{equation}
\begin{equation}\label{PT2}
[H,J_-]=-2\sqrt{H}J_--J_-,
\end{equation}
\begin{equation}\label{PT3}
[J_-,J_+]=2\sqrt{H}(2H+3\sqrt{H}+\mathbbm{1}).
\end{equation}
Note that the difference between the GHA \cite{Curado2} and the introduced generalized su(1,1) algebra of the P\"{o}schl-Teller potential is the commutator between the creation and the annihilation operators given in (\ref{PT3}).\\ The Fock space representation of the generalized su(1,1) can be obtained by considering an eigenvector $\ket{n}$ of $H$. From (\ref{j++}) -(\ref{j--}) and (\ref{varepsi}), the action of $H$, $J_+$ and $J_-$ on an eigenvector $\ket{n}$ can be given by 
\begin{equation}\label{Hpt}
H\ket{n}=(n+\nu+1)^2\ket{n},\quad n=0,1,2,\dots,
\end{equation}
	\begin{widetext} 
		\begin{equation}\label{J+Pt}
J_+\ket{n}=\sqrt{(n+1)(n+2\nu+3)((n+2+\nu)^2+(\nu+1)^2-1)}\ket{n+1},
	\end{equation}
 \begin{equation}\label{J-Pt}
	J_-\ket{n}=\sqrt{n(n+2\nu+2)((n+1+\nu)^2+(\nu+1)^2-1)}\ket{n-1}.
	\end{equation}
\end{widetext}
Note that $J_-\ket{0}=0$. The physical realization of the generalized su(1,1)  generators can be easily constructed. By using (\ref{Hpt})-(\ref{J-Pt}) and (\ref{A2})-(\ref{A3}), the algebra generators in terms of the position and differential operators can be given by 
	\begin{equation}
H=-\dfrac{L^2}{\pi^2}\dfrac{d^2}{dx^2}+\dfrac{\nu(\nu+1)}{\sin^2(\pi x/L)},
\end{equation}
\begin{widetext}
\begin{equation}\label{ladder1}
J_{+}=\left\{\frac{L}{\pi}\sin(\frac{\pi x}{L})\dfrac{d}{dx}+\cos(\frac{\pi}{L}x)(N+\nu+1)\right\}k(N),
\end{equation}
and
\begin{equation}\label{ladder2}
J_{-}=k(N)\left\{-\frac{L}{\pi}\dfrac{d}{dx}\sin(\frac{\pi x}{L})+(N+\nu+1)\cos(\frac{\pi}{L}x)\right\},
\end{equation}
where
\begin{equation}
k(N)=\sqrt{\dfrac{(N+\nu+2)(N+2\nu+3)[(N+2+\nu)^2+(\nu+1)^2-1]}{(N+\nu+1)(N+2\nu+2)}}.
\end{equation}
\end{widetext}
By using (\ref{ladder1})-(\ref{ladder2}) and (\ref{A2})-(\ref{A3}), we can easily show that $J_+\psi_n(x)=\sqrt{(\varepsilon_{n+1}-\varepsilon_{0})(\varepsilon_{n+1}+\varepsilon_{0}-1)}\psi_{n+1}(x)$ and $J_-\psi_n(x)=\sqrt{(\varepsilon_{n}-\varepsilon_{0})(\varepsilon_{n}+\varepsilon_{0}-1)}\psi_{n-1}(x)$  where $\varepsilon_n$ is given in (\ref{varepsi}). This shows how the physical system can be described algebraically by the generalized su(1,1) and showing the correspondence between the algebraic approach and the wave function approach. 
For the P\"{o}schl-Teller potential, the operators $B$ and $B^\dagger$ given in (\ref{B}) and satisfying the harmonic oscillator commutation relations are given by 
\begin{widetext}
\begin{equation}
B=\sqrt{\dfrac{1}{(N+2\nu+3)((N+2+\nu)^2+(\nu+1)^2-1)}}J_-,
\end{equation}
\begin{equation}
B^\dagger=J_+\sqrt{\dfrac{1}{(N+2\nu+3)((N+2+\nu)^2+(\nu+1)^2-1)}}.
\end{equation}
\end{widetext}
	\section{Coherent states for the   P\"{o}schl-Teller potential with $\nu=0$}\label{section4}
	\subsection{Generalized su(1,1) nonlinear Coherent states }
	In this section, we aim to construct the Barut-Girardello coherent states \cite{Barut} for  the   P\"{o}schl-Teller potential associated with the generalized su(1,1) algebra. Let us first recall the minimal set of conditions necessary to construct a Klauder's coherent state. A state $\ket{z}$ is a Klauder's coherent state if and only if it satisfies the following conditions \\ 
 	i)normalization $\braket{z|z}=1$ , \\
 	ii) continuity in the label,
 	\begin{equation}\label{continuitycondition}
 	|\left\vert z \right\rangle - \left\vert z' \right\rangle|\longrightarrow 0 \quad \text{when } \quad | z- z'|\longrightarrow 0,
 	\end{equation}
 	iii) completeness or resolution of the identity
 	\begin{widetext} \begin{equation}\label{completeness}
 	\int\int d^2z W(|z|^2)\left\vert z \right\rangle\left\langle z \right\vert=\sum_{n=0}^{\infty}\ket{n}\bra{n}=\mathbbm{1},
 	\end{equation}
 \end{widetext}
 where $W(|z|^2)$ is a positive function called the weight function. For the sake of simplification, we will take $\nu=0$. Thus, the spectrum (\ref{varepsi}) now has the same form as a free particle in a square well potential. In the literature, several nonlinear coherent states for P\"{o}schl-Teller have been constructed \cite{Curado2,Chen_1998,Bergeron,Banerji, Kinani,Cruz,ncheyta,Hounkonnou}. In \cite{Curado2}, the nonlinear coherent state $\ket{z}_{NL1}$ associated with GHA is defined by 
	\begin{equation}
	A\ket{z}_{NL1}=z\ket{z}_{NL1}.
	\end{equation} 
Remembering that $N_n=\varepsilon_{n+1}-\varepsilon_{0}$. For the  P\"{o}schl-Teller potential with $\nu=0$, $N_n=\sqrt{(n+1)(n+3)}$ and $\ket{z}_{NL1}$ reads
	\begin{equation}\label{NL1}
	\ket{z}_{NL1}=N_1(|z|^2)\sum_{n=0}^{\infty}\dfrac{z^n}{N_{n-1}!}\ket{n},
	\end{equation}
where $N_{n-1}!=N_0N_1\dots N_{n-1}$ and by definition $N_{-1}=1$	and $z$ is a complex number. This vector satisfies the normalization, continuity in the label and resolution of unity \cite{Curado2,Klau,PhysRevA.64.013817}. The normalization factor given in (\ref{NL1}) is
	\begin{equation}
(N_1(|z|^2))^2=\dfrac{|z|^2}{2I_2(2|z|)},
	\end{equation}
	where $I_2$ is the modified Bessel function of the second kind \cite{Gradshteyn}.\\
	The linear coherent state associated with  P\"{o}schl-Teller potential has been constructed in \cite{Curado2}. It is given by 
	\begin{equation}\label{L1}
\ket{z}_{L}=\text{e}^{-|z|^2/2}\sum_{n=0}^{\infty}\dfrac{z^n}{\sqrt{n!}}\ket{n}.
\end{equation}
where $\braket{x|n}=\psi_{n}(x)$ is the wave function (\ref{psipt}) of the  P\"{o}schl-Teller system. The state (\ref{L1}) is defined as an eigenstate of the generalized harmonic oscillator annihilation operator \cite{Curado2} and can be constructed, as a particular case, by using the approach provided in \cite{PhysRevLett.72.1447}.\\ Now, we construct the generalized su(1,1) algebra nonlinear coherent state. Let us consider a vector $\ket{z}_{NL2}$ satisfying  \cite{Hassouni2}
\begin{equation}
J_-\ket{z}_{NL2}=z\ket{z}_{NL2}
\end{equation}
Let $\mathcal{N}_n=\sqrt{(\varepsilon_{n+1}-\varepsilon_{0})(\varepsilon_{n+1}+\varepsilon_{0}-1)}$, the vector $\ket{z}_{NL2}$  can be given by
\begin{equation}\label{akher}
\ket{z}_{NL2}=N_{2}(|z|^2)\sum_{n=0}^{\infty}\dfrac{z^n}{\mathcal{N}_{n-1}!}\ket{n}.
\end{equation}
where $N_{2}(|z|^2)$ is a normalization factor,  $\mathcal{N}_{n-1}!=\mathcal{N}_0\mathcal{N}_1\dots\mathcal{N}_{n-1}$ and $\mathcal{N}_{-1}!:=0$.\\ By using (\ref{J-Pt}) and (\ref{akher}), for the P\"{o}schl-Teller potential (with $\nu=0$), $\ket{z}_{NL2}$ can be written as 
\begin{equation}\label{CSSU11}
\ket{z}_{NL2}=N_{2}(|z|^2)\sum_{n=0}^{\infty}\dfrac{\sqrt{2}z^n}{(n+1)!\sqrt{n!(n+2)!}}\ket{n}.
\end{equation}
We require that $\braket{z|z}_{NL2}=1$. Then, the normalization factor $N_{2}(|z^2|)$ reads 
\begin{equation}
N_{2}(|z|^2)=\frac{1}{\sqrt{\, _0F_3\left(;2,2,3;|z|^2\right)}},
\end{equation}
where $_0F_3\left(;2,2,3;|z|^2\right)$ is  the generalized hyper-geometric function. The normalization function $N_{2}(|z|^2)$ is defined for all $|z|\in [0,\infty[$. Thus, $z$ in (\ref{CSSU11}) belongs to the whole complex plan. The continuity condition (\ref{continuitycondition}) is simply verified. We  give now the appropriate weight function satisfying the completeness relation (\ref{completeness}). Let $z=r\text{e}^{i\theta}$ where $0\leq r<\infty$ and $0\leq \theta\leq 2\pi$. Substituting (\ref{CSSU11}) in (\ref{completeness}), it follows that
\begin{widetext}
\begin{multline}\label{Multi}
\sum_{n,n^{'}=0}^\infty\left\{ \dfrac{1}{2}\dfrac{1}{(n+1)!(n^{'}+1)!\sqrt{n!n^{'}!(n+2)!(n^{'}+2)!}}\int_{0}^{\infty}W(r^2)N^2_2(r^2)2r^{n+n^{'}}d (r^2)\int_{0}^{2\pi}\text{e}^{i\theta(n-n^{'})}d\theta\right\}\\\ket{n}\bra{n^{'}}=\sum_{n=0}^\infty\left\{\dfrac{2\pi}{((n+1)!)^2n!(n+2)!}\int_{0}^\infty x^n W(x)N^2_2(x)dx\right\}\ket{n}\bra{n}=\mathbbm{1},\quad (x=r^2),\hspace{2cm} 
\end{multline}
\end{widetext}
where we have used the fact that $\int_{0}^{2\pi}\text{e}^{i(n-n^{'})\theta}d\theta=2\pi\delta_{n,n^{'}}$. Remembering that  $\sum_{n=0}^\infty\ket{n}\bra{n}=\mathbbm{1}$, the completeness condition (\ref{Multi}) reduces to

\begin{equation}\label{integralequ}
\int_{0}^\infty x^ng(x)=\dfrac{((n+1)!)^2n!(n+2)!}{2},
\end{equation}
\\

where $g(x)=\pi N^2_2(x)W(x)$. 
Now, we apply the methods recalled in (\ref{appen}) to solve the moment problem (\ref{integralequ}). We have 
\begin{equation}
2\int_0^\infty dx x^n x K_0(2\sqrt{x})=((n+1)!)^2,
\end{equation}
and 
\begin{equation}
2\int_0^\infty dx x^n x K_2(2\sqrt{x})=n!(n+2)!.
\end{equation}
where $K_n(x)$ is the modified Bessel function of the second kind. Thus, By using (\ref{1})-(\ref{2}), the moment problem (\ref{integralequ}) can be solved by 
\\

\begin{figure}[h]
	\centering
	\includegraphics[width=7cm]{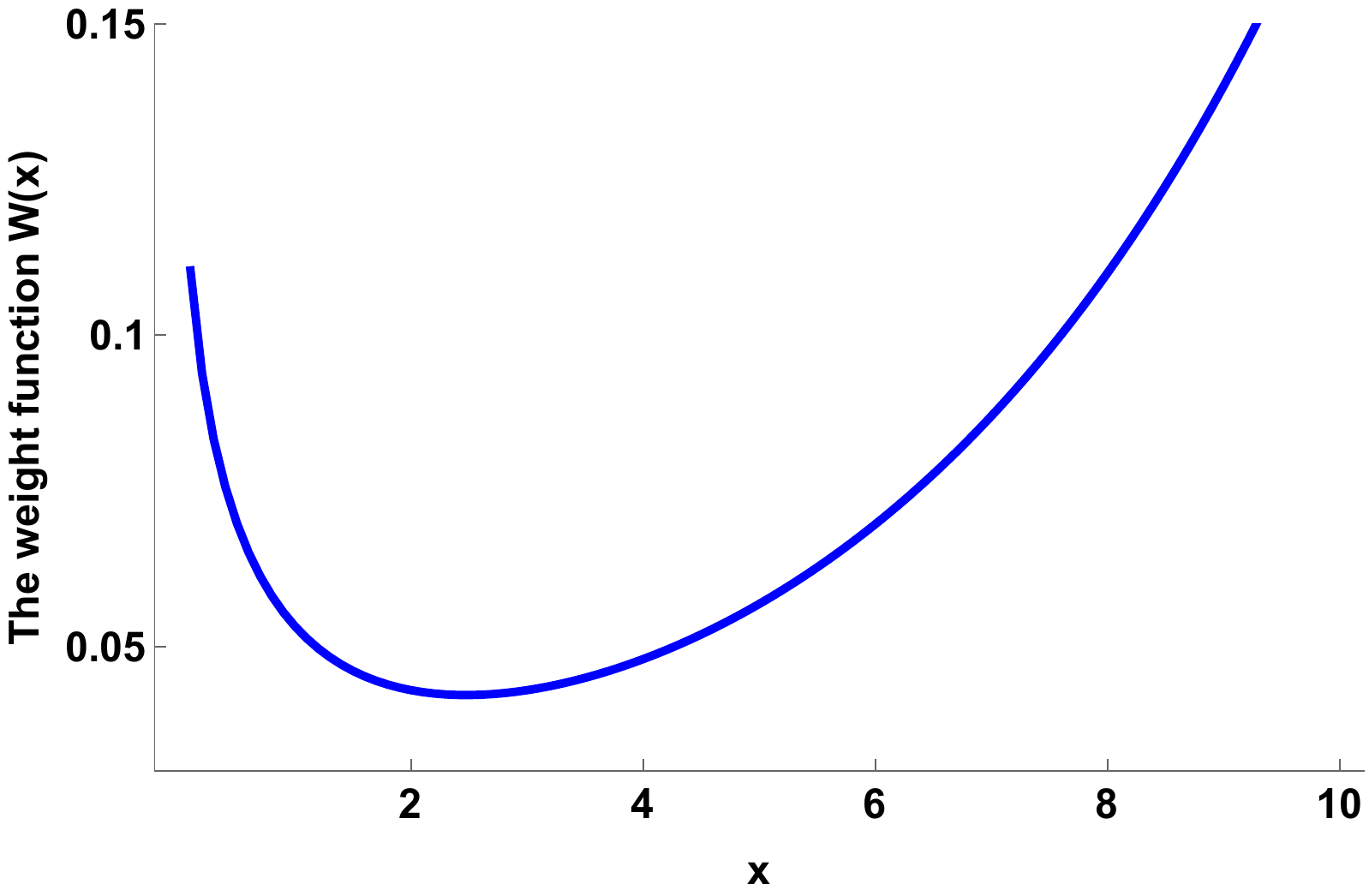}
	\caption{The behavior of the weight function for the P\"oschl-Teller potential in terms of $x=|z|^2.$} 
	\label{fig1}
\end{figure}

\begin{widetext}
\begin{equation}
g(x)=\int_{0}^\infty2xK_0(2\sqrt{x/t})K_2(2\sqrt{t})\dfrac{dt}{t},
\end{equation}
Finally, the weight function satisfying the completeness condition (\ref{completeness}), in this case, is 
\begin{equation}
	W(x)=\dfrac{1}{\pi N^2_2(x)}\int_{0}^\infty2xK_0(2\sqrt{x/t})K_2(2\sqrt{t})\dfrac{dt}{t}.
	\end{equation}
\end{widetext}
The behavior of the $W(x)$ is shown in figure (\ref{fig1}).  We can see that $W(x)$ is a positive function.\\
Consequently, the vector $ \ket{z}_{NL2}$ is a Klauder coherent state. Hence, two kinds of nonlinear coherent states can be constructed for the P\"{o}schl-Teller potential, one is associated with GHA and the second is associated with the generalized su(1,1) algebra. Note that the coherent states constructed here differs from the ones constructed in \cite{ Kinani,Cruz,Hounkonnou} where the factorization of the Hamiltonian of the P\"oschl-Teller potential in terms of the ladder operators is needed. Here, we have constructed the generalized su(1,1) nonlinear coherent states for the P\"oschl-Teller potential in an algebraic manner by using the methods followed in \cite{Hassouni2,Curado2}. 
\subsection{Time evolution of nonlinear coherent states of the   P\"{o}schl-Teller potential } 
We compare now the behaviors of the time evolution of the uncertainty relation for both nonlinear coherent states. For simplification of the calculations, from now on, we will consider that the Fock space starts from $n=1$ instead of $n=0$. Then,  the GHA nonlinear coherent state (\ref{NL1}) can be now written as 
	\begin{equation}\label{GHAPTCS}
\ket{z}_{NL1}=N_1(|z|^2)\sum_{n=1}^{\infty}\dfrac{z^{n-1}}{\prod_{i=2}^{n}(i^2-1)}\ket{n},
\end{equation}
while the generalized su(1,1) nonlinear coherent state introduced in (\ref{CSSU11}) becomes
\begin{equation}\label{nl1}
\ket{z}_{NL2}=N_{2}(|z|^2)\sum_{n=1}^{\infty}\dfrac{\sqrt{2}z^{n-1}}{n!\sqrt{(n-1)!(n+1)!}}\ket{n}.
\end{equation}
The time evolution of a given state can be obtained by the action of the evolution operator 
\begin{equation}
U(t)=\text{e}^{-iHt/\hbar}.
\end{equation}
Let $\dfrac{\hbar^2\pi^2}{2mL^2}=1$. This means that we choose a particular P\"{o}schl-Teller system. Then, at $t$ we have 
	\begin{equation}\label{GHAPTCS2}
\ket{z,t}_{NL1}=N_1(|z|^2)\sum_{n=1}^{\infty}\dfrac{z^{n-1} \text{e}^{-in^2t/\hbar}}{\prod_{i=2}^{n}(i^2-1)}\ket{n},
\end{equation}
and
\begin{equation}\label{XNL1}
\ket{z,t}_{NL2}=N_{2}(|z|^2)\sum_{n=1}^{\infty}\dfrac{\sqrt{2}z^{n-1}\text{e}^{-in^2t/\hbar}}{n!\sqrt{(n-1)!(n+1)!}}\ket{n}.
\end{equation}
The time evolution of $X(t)$, $P(t)$, $X^2(t)$ and $P^2(t)$ on the state (\ref{GHAPTCS2})  can be easily obtained. They were calculated in \cite{Curado2} and can be written as 
\begin{widetext}
	\begin{multline}\label{XtPT}
	\braket{X(t)}_{NL1}=\dfrac{L}{\sqrt{2}}{}\braket{z,t|B+B^\dagger|z,t}_{NL1}\\
	=\dfrac{\sqrt{2}Lr^2}{I_2(2r)}\sum_{n=1}^\infty\dfrac{r^{2n-1}\cos((2n+1)t/\hbar-\theta)}{\sqrt{n+2}\Gamma(n)\Gamma(n+2)},\hspace{9.2cm}
	\end{multline}
	\begin{multline}
	\braket{P(t)}_{NL1}=\dfrac{i\hbar}{\sqrt{2}L}\braket{z,t|B^\dagger-B|z,t}_{NL1}\\
	=-\dfrac{\sqrt{2}\hbar r^2}{LI_2(2r)}\sum_{n=1}^\infty\dfrac{r^{2n-1}\sin((2n+1)t/\hbar-\theta)}{\sqrt{n+2}\Gamma(n)\Gamma(n+2)},\hspace{8.8cm}
	\end{multline}
	
	\begin{multline}
	\braket{X^2(t)}_{NL1}=\dfrac{L^2}{2}\braket{z,t|B^2+(B^\dagger)^2+B^\dagger B+BB^\dagger|z,t}_{NL1}\\
	=\dfrac{L^2r^2}{I_2(2r)}\left(\sum_{n=1}^\infty\dfrac{r^{2n}\sqrt{n+1}\cos((4n+4)t/\hbar-2\theta)}{\sqrt{(n+2)(n^2+4n+3)}\Gamma(n)\Gamma(n+2)}+\sum_{n=1}^\infty\dfrac{(n-1)r^{2n-2})}{\Gamma(n)\Gamma(n+2)}\right)+\dfrac{L^2}{2},\hspace{3cm}
	\end{multline}
	\begin{multline}
\braket{P^2(t)}_{NL1}=-\dfrac{\hbar^2}{2L^2}\braket{z,t|B^2+(B^\dagger)^2-B^\dagger B-BB^\dagger|z,t}_{NL1}\\
=-\dfrac{\hbar^2r^2}{L^2I_2(2r)}\left(\sum_{n=1}^\infty\dfrac{r^{2n}\sqrt{n+1}\cos((4n+4)t/\hbar-2\theta)}{\sqrt{(n+2)(n^2+4n+3)}\Gamma(n)\Gamma(n+2)}-\sum_{n=1}^\infty\dfrac{(n-1)r^{2n-2})}{\Gamma(n)\Gamma(n+2)}\right)+\dfrac{\hbar^2}{2L^2}.\hspace{2.1cm}
\end{multline}

We repeat the same calculations for the nonlinear coherent state (\ref{XNL1}), we find that 
\begin{equation}
\braket{X(t)}_{NL2}=2\sqrt{2}LN_2^2(r^2)\sum_{n=1}^\infty\dfrac{r^{2n-1}\sqrt{n}\cos((2n+1)t/\hbar-\theta)}{n!(n+1)!\sqrt{(n-1)!n!(n+1)!(n+2)!}},\hspace{4.6cm}
\end{equation}
\begin{equation}
	\braket{P(t)}_{NL2}=\dfrac{-2\sqrt{2}\hbar N^2_2(r^2)}{L}\sum_{n=1}^\infty\dfrac{r^{2n-1}\sqrt{n}\sin((2n+1)t/\hbar-\theta)}{n!(n+1)!\sqrt{(n-1)!n!(n+1)!(n+2)!}},\hspace{4.6cm}
	\end{equation}
	\begin{equation}
\braket{X^2(t)}_{NL2}
=2L^2N_2^2(r^2)\left(\sum_{n=1}^\infty\dfrac{r^{2n}\sqrt{n(n+1)}\cos((4n+4)t/\hbar-2\theta)}{n!(n+2)!(n+1)!\sqrt{(n-1)!(n+3)!}}+\sum_{n=1}^\infty\dfrac{(n-1)r^{2n-2}}{(n!)^2(n-1)!(n+1)!}\right)+\dfrac{L^2}{2},
\end{equation}
	\begin{equation}\label{XNL2}
\braket{P^2(t)}_{NL2}=\dfrac{-2\hbar^2N^2_2(r^2)}{L^2}\left(\sum_{n=1}^\infty\dfrac{r^{2n}\sqrt{n(n+1)}\cos((4n+4)t/\hbar-2\theta)}{n!(n+2)!(n+1)!\sqrt{(n-1)!(n+3)!}}-\sum_{n=1}^\infty\dfrac{(n-1)r^{2n-2}}{(n!)^2(n-1)!(n+1)!}\right)+\dfrac{\hbar^2}{2L^2}.
\end{equation}	
\end{widetext}

\begin{figure}[t]
  	\centering
  	\includegraphics[width=7cm]{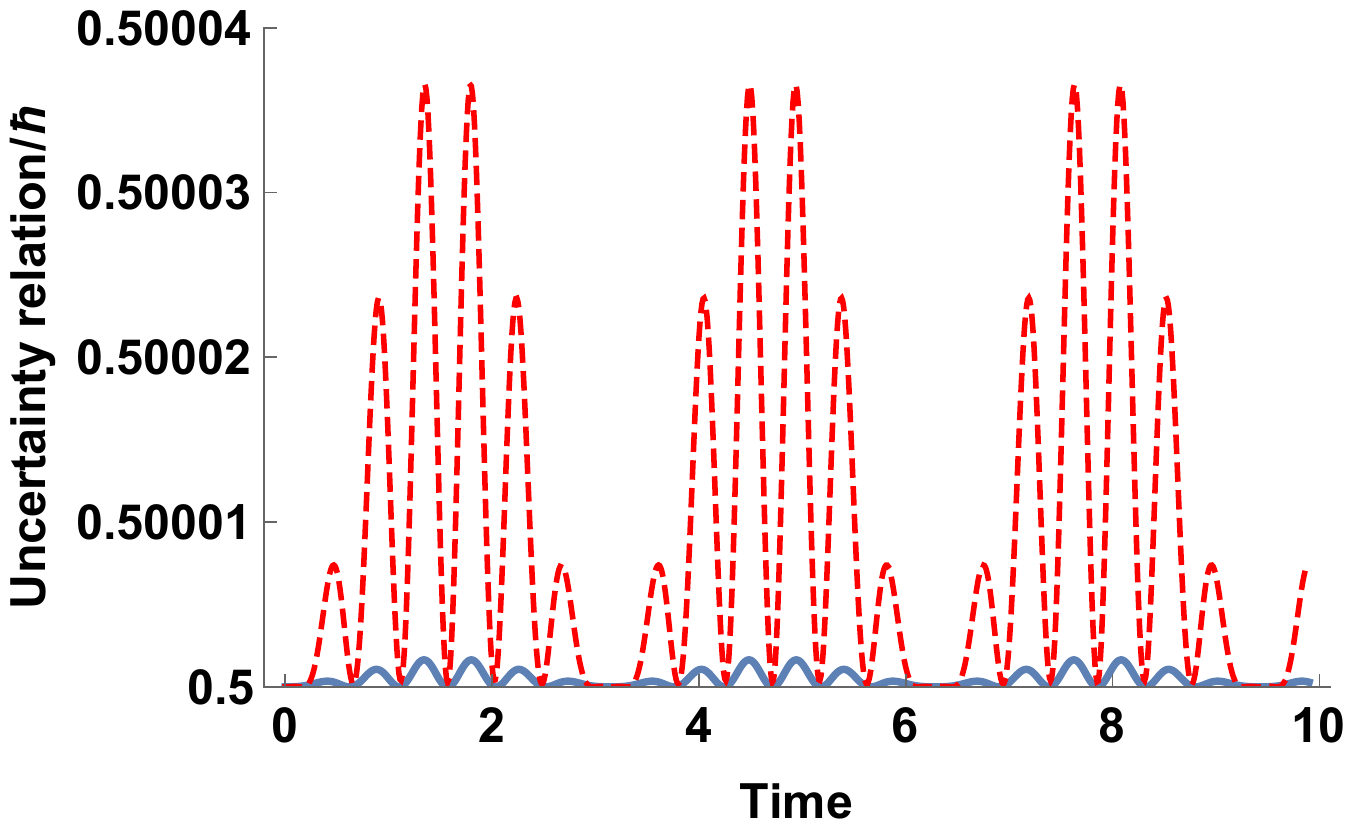}
  	\caption{  Time evolution of the uncertainty relation for $r=0.1$, $\theta=0$ and $\hbar=1$ for GHA nonlinear coherent state (dashed line) and generalized su(1,1) nonlinear coherent state(continuous line)} 
  	\label{fig2}
  \end{figure}

\begin{figure}[h]
	\centering
	\includegraphics[width=7cm]{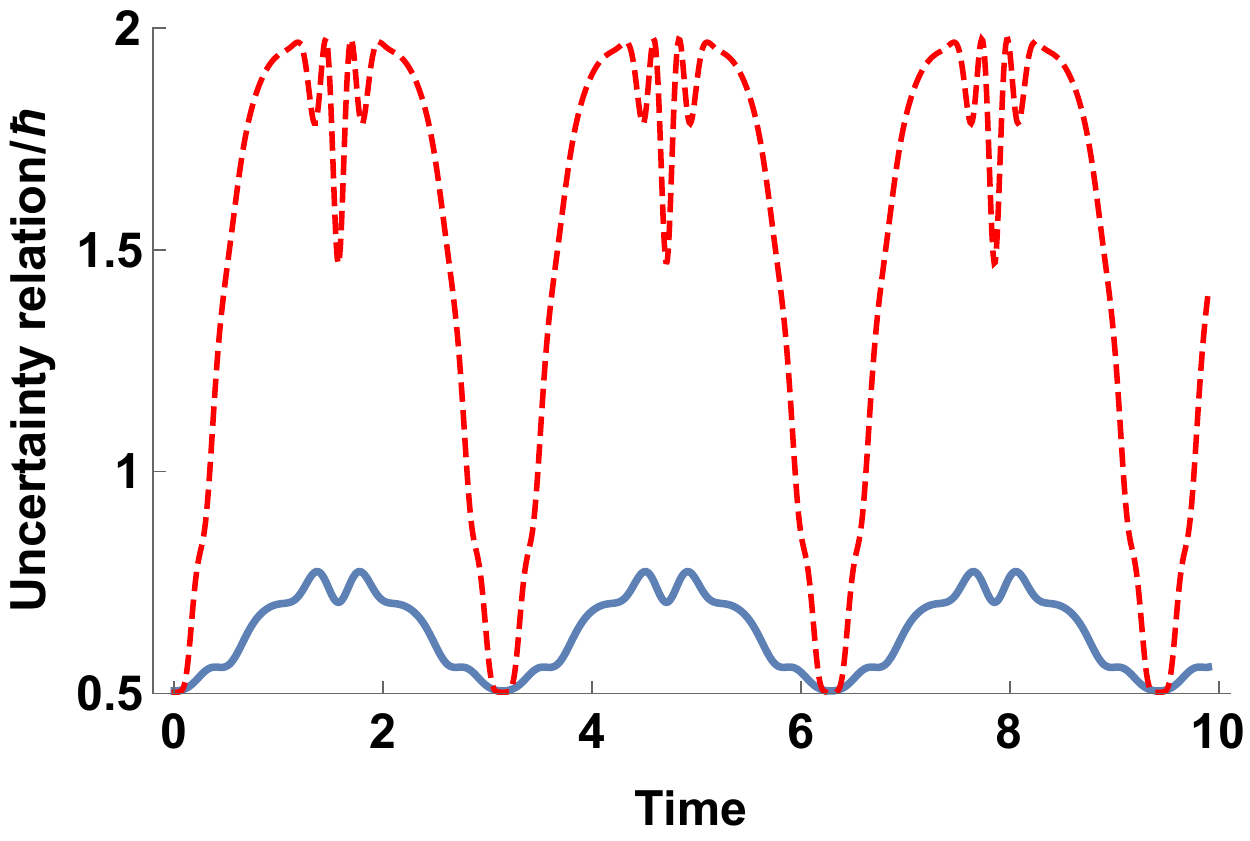}
	\caption{  Time evolution of the uncertainty relation for $r=2.5$, $\theta=0$ and $\hbar=1$ for GHA nonlinear coherent state (dashed line) and generalized su(1,1) nonlinear coherent state (continuous line)} 
	\label{fig3}
\end{figure}

\begin{figure}[h]
	\centering
	\includegraphics[width=7cm]{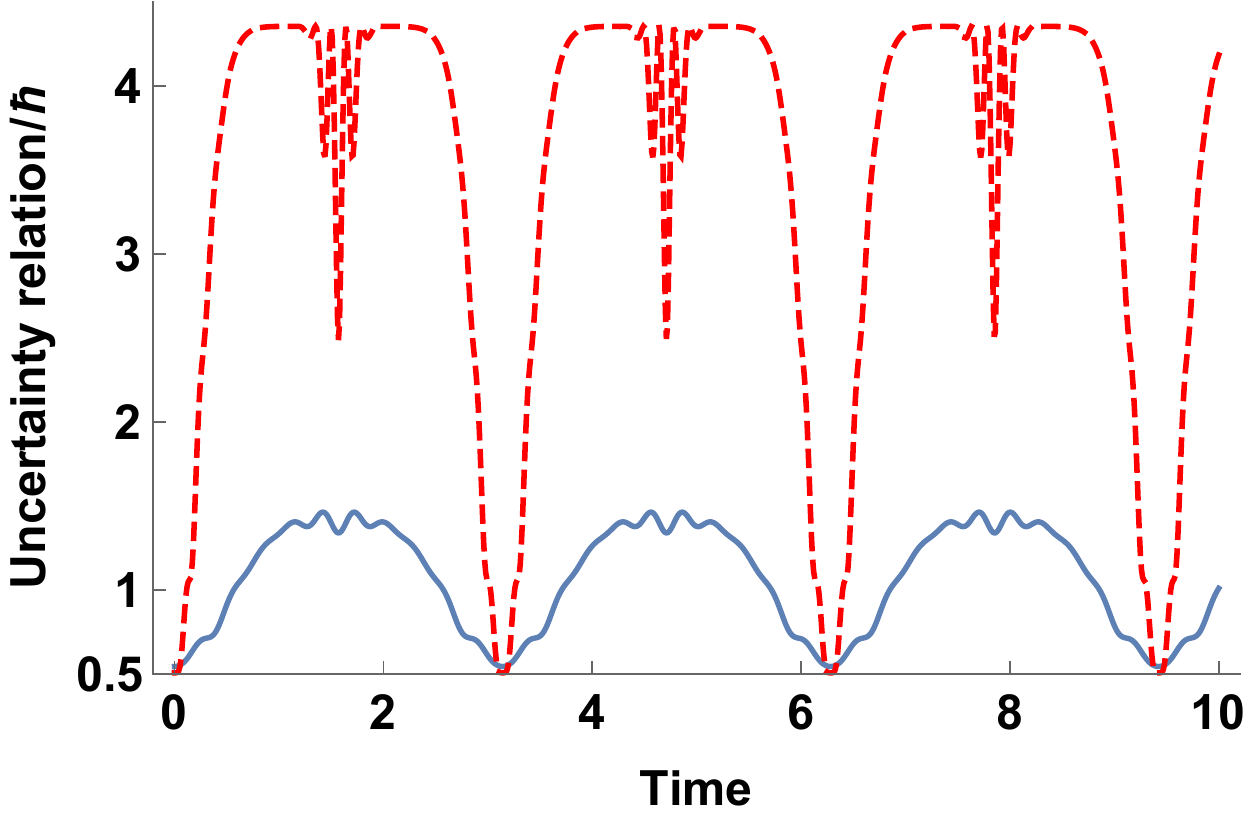}
	\caption{  Time evolution of the uncertainty relation for $r=5$, $\theta=0$ and $\hbar=1$ for GHA nonlinear coherent state (dashed line) and generalized su(1,1) nonlinear coherent state( continuous line)} 
	\label{fig4}
\end{figure}

The time evolution of the uncertainty relation is given by
\begin{multline}\label{timeevolution}
\Delta X(t)\Delta P(t)=\sqrt{\braket{X^2(t)}-\braket{X(t)}^2}\times\\\sqrt{\braket{P^2(t)}-\braket{P(t)}^2}.\hspace{2.1cm}
\end{multline}
It can be easily obtained for both nonlinear coherent states  from (\ref{XtPT})-(\ref{XNL2}).
\\ In \cite{Curado2}, the time evolution (\ref{timeevolution})  of linear coherent states and GHA nonlinear coherent states were calculated and it was shown that the GHA nonlinear ones are the more localized. The time evolution of coherent states has been studied in several works \cite{Curado2,Mesoscopic}. Particularely, in \cite{Mesoscopic} the authors studied the time evolution and  the  phase-space dynamics of su(2) Morse potential coherent states by using the Wigner function.\\
Let us now compare the behavior of the time evolution of the uncertainty relation of the  generalized su(1,1) coherent states and that of GHA  ones. 
  In Figures (\ref{fig2})-(\ref{fig4}) the time evolution (\ref{timeevolution}) of both nonlinear coherent states are shown for $r=0.1$, $r=5$ and $r=10$; $\theta=0$ and $\hbar=1$.
 Analyzing these  figures, we can see that for both
  nonlinear coherent states, the uncertainty relations oscillates between $0.5\hbar$ and maximum values. We see also that the uncertainties of both coherent states are equal in few points and that the uncertainty of generalized su(1,1) coherent
 states is always smaller than that of GHA coherent states.  We can conclude that our nonlinear coherent states are more localized than  GHA ones.  We can see also in figures (\ref{fig2})-(\ref{fig4}) that the maximum uncertainty for both nonlinear coherent states increases with increasing $r$ which indicates that the uncertainty approaches to $0.5\hbar$  for small values of $r$.  Since the generalized su(1,1) nonlinear coherent states for the P\"oschl-Teller potential are more localized. Then, they  take to wave packet more closer  to the classical trajectory than that of GHA nonlinear coheren states.

\subsection{Phase-space trajectories for $X$ and $P$ } 
\begin{figure}[h]
	\centering
	\includegraphics[width=7cm]{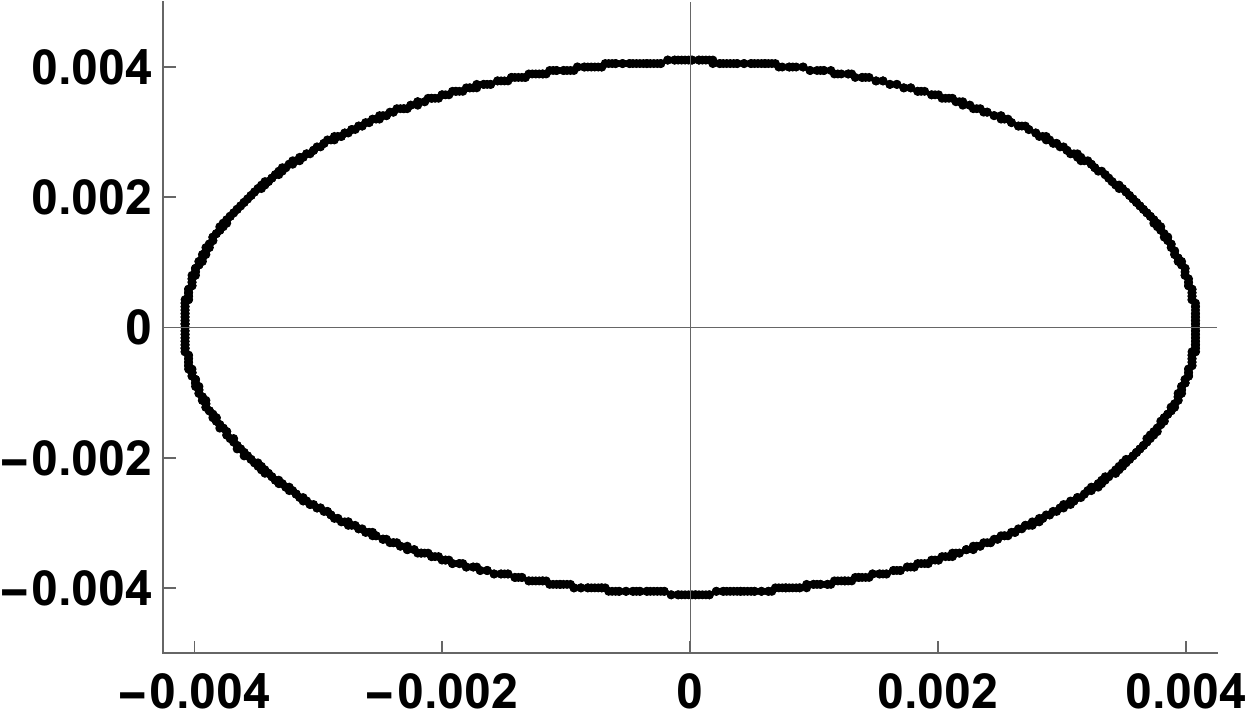}
	\caption{$X$ and $P/\hbar$ phase-space trajectory of Generalized su(1,1) P\"oschl-Teller coherent state with $r=0.01$.} 
	\label{fig6}
\end{figure}

\begin{figure}[t]
	\centering
	\includegraphics[width=7cm]{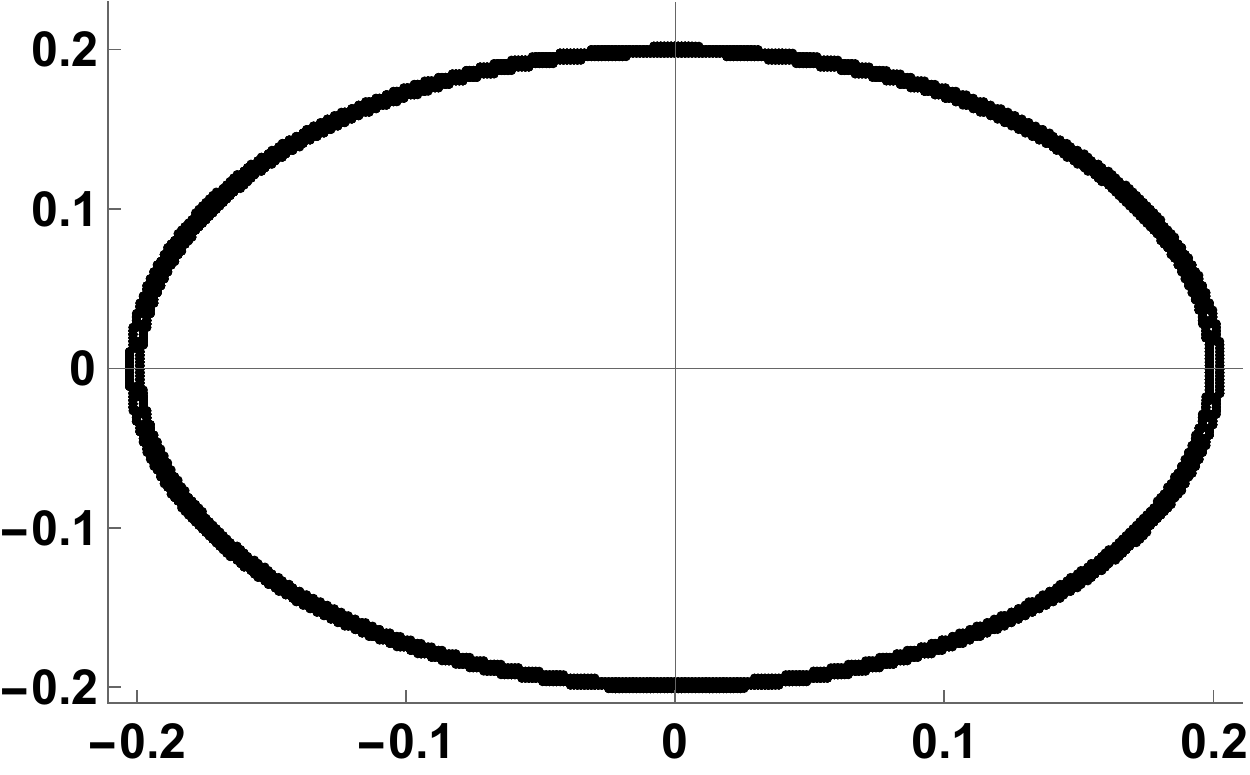}
	\caption{$X$ and $P/\hbar$ phase-space trajectory with of  of Generalized su(1,1) P\"oschl-Teller coherent state with $r=0.5$. } 
	\label{fig7}
\end{figure}

We now analyze the phase space trajectories of $ (X,P/\hbar)$ of generalized su(1,1) coherent states.
In the figures (\ref{fig6})-(\ref{fig7}), we show $ (X,P/\hbar)$ for $r=0.01$ and $r=0.5$, respectively. In these two graphs, we have taken $L=\hbar=1$.  We can see that the form of the phase space  trajectory is an ellipse for $r=0.01$ and $r=0.5$. The phase space trajectory of  GHA nonlinear coherent states were obtained in \cite{Curado2}. It is an ellipse only for very small values of $r$\cite{Curado2}.

\section{Final comment}\label{section5}
In this paper, we have introduced a generalization structure of the su(1,1) algebra which depends on a given function of one hermitian operator, called the characteristic function of the algebra. The irreducible representation of the algebra generators has been given and it has been shown that a hidden symmetry is  presented in the sequence of eigenvalues of the hermitian operator. The generalized su(1,1) algebra constructed here bears some analogy with the GHA and since this latter can be applied to  physical system whose two successive energy eigenvalues are connected through the characteristic function of the GHA, we believe that these physical systems can also be described by the  generalized su(1,1) algebra and the characteristic function of this algebra is the same as that of the GHA. We have examined the P\"{o}schl-Teller potential and have expressed the generators of the constructed generalized su(1,1) algebra in terms of  position and differential operators. Then, we have constructed the  pair of canonically conjugate operators satisfying  the bosonic relations of the harmonic oscillator algebra. Furthermore, we have constructed the Barut-Girardello coherent states for P\"{o}schl-Teller potential and compare the behavior of the time evolution of the uncertainty relation of the constructed nonlinear coherent states and that of  GHA coherent states. The generalized su(1,1) coherent states are more localized. Thence, we have seen that  the  uncertainty almost equal $\hbar/2$ for small values of radius of convergence of the constructed nonlinear coherent states. An interesting result is that uncertainty oscillates between a minimum value ( which is $\hbar/2$) and  maximum values close to $\hbar/2$ compared with GHA nonlinear coherent states whose  uncertainty oscillates between $\hbar/2$ and  maximum values which are so large than $\hbar/2$.
\section*{Acknowledgement} 
	This work is partially supported by the ICTP through AF-14.
\begin{widetext}
\appendix
\section{Gegenbauer  recurrence  formulae}
The energy eigenfunctions associated with Schr\"{o}dinger equation (\ref{PT}) are given by 
\begin{equation}
\psi_{n}(x)=c_{n}(\nu)\sin^{\nu+1}(\pi x/L)C_{n}^{\nu+1}(\cos(\pi x/L)),\quad\text{where}\quad  c_{n}(\nu)=\Gamma(\nu+1)\dfrac{2^{\nu+1/2}}{\sqrt{L}}\sqrt{\dfrac{n!(n+\nu+1)}{\Gamma(n+2\nu+2)}},
\end{equation}
and $C_{n}^{\nu+1}(\cos(\pi x/L))$ are the Gegenbauer polynomials.\\
By using the recurrence formulae satisfied by  Gegenbauer polynomials \cite{Wilhelm}, one can show that
\begin{equation}\label{A2}
\cos(\pi x/L)\psi_{n}(x)=\dfrac{1}{2\sqrt{n+\nu+1}}\left(\sqrt{\dfrac{n(n+2\nu+1)}{n+\nu}}\psi_{n-1}(x)+ \sqrt{\dfrac{(n+1)(n+2\nu+2)}{n+\nu+2}}\psi_{n+1}(x)\right)
\end{equation}
and that 
\begin{equation}\label{A3}
-\dfrac{L}{\pi}\sin(\pi x/L)\dfrac{d\psi_{n}(x)}{dx}=\dfrac{\sqrt{n+\nu+1}}{2}\left(\sqrt{\dfrac{n(n+2\nu+1)}{n+\nu}}\psi_{n-1}(x)- \sqrt{\dfrac{(n+1)(n+2\nu+2)}{n+\nu+2}}\psi_{n+1}(x)\right)
\end{equation}
\section{Mellin transform}\label{appen}
Let $l(x)$ be an analytical function. The Mellin transform \cite{fitouhi,Oberhettingerr,Wilhelm}, $l^*(s)$ of the function $l(x)$ is defined by
\begin{equation}
l^*(s):=\int_{0}^{\infty}l(x)x^{s-1}dx,
\end{equation} 
where $ s$ is a complex variable. The function $l(x)$ is the inverse Mellin transform of the function $l(s)$ and it reads as
\begin{equation}\label{MTI}
l(x)=\dfrac{1}{2\pi i}\int_{c-i\infty}^{c+i\infty}l^*(s)x^{-s}ds,
\end{equation}
where $c$ is a complex number and $i^2=-1$. Let $l(x)$ and $h(x)$ be two functions, the Mellin convolution reads 
\begin{equation}\label{Convolution}
x^a\int_{0}^\infty t^b l(\dfrac{x}{t})h(t)dt=\dfrac{1}{2\pi i}\int_{-i\infty}^{-i\infty}l^*(s+a)h^*(s+a+b+1)x^{-s}ds,
\end{equation}
where $a$, $b$ are arbitrary complexes. From (\ref{Convolution}), it follows that if 
\begin{equation}
l_n!=\int_{0}^\infty x^nl(x)dx,\quad  \text{and}\quad h_n!=\int_{0}^\infty x^nh(x)dx,
\end{equation}
where $l_n!=l_nl_{n-1}\dots l_0$ and $h_n!=h_nh_{n-1}\dots h_0$. Then, the weight function defined by 
\begin{equation}\label{1}
a(x)=\int_{0}^\infty l(x/t)h(t)\dfrac{dt}{t},
\end{equation}
is the solution of the moment problem 
\begin{equation}\label{2}
l_n!h_n!=\int_{0}^\infty x^n a(x) dx.
\end{equation}
\end{widetext}

\end{document}